\documentclass[sigconf,nonacm=true]{acmart}

\usepackage{latexsym}
\usepackage{amsmath}
\usepackage{amsthm}
\usepackage{booktabs}
\usepackage{enumitem}
\usepackage{graphicx}
\usepackage{color}
\usepackage{tcolorbox}
\usepackage{tabularx}
\usepackage{multirow}
\DeclareUnicodeCharacter{FF0C}{,}

\newcommand{\etal}{\textit{et al.}}

\begin{document}
\title{Large Language Model-driven Multi-Agent Simulation for News Diffusion Under Different Network Structures}

\author{Xinyi Li}
\affiliation{%
\institution{Northwestern University}
\state{Illinois}
\country{United States}
}

\author{Yu Xu}
\affiliation{%
\institution{Northwestern University}
\state{Illinois}
\country{United States}}

\author{Yongfeng Zhang}
\affiliation{%
\institution{Rutgers University}
\state{New Jersey}
\country{United States}}

\author{Edward C. Malthouse}
\affiliation{%
\institution{Northwestern University}
\state{Illinois}
\country{United States}}

\begin{abstract}
The proliferation of fake news in the digital age has raised critical concerns, particularly regarding its impact on societal trust and democratic processes. Diverging from conventional agent-based simulation approaches, this work introduces an innovative approach by employing a large language model (LLM)-driven multi-agent simulation to replicate complex interactions within information ecosystems. We investigate key factors that facilitate news propagation, such as agent personalities and network structures, while also evaluating strategies to combat misinformation. Through simulations across varying network structures, we demonstrate the potential of LLM-based agents in modeling the dynamics of misinformation spread, validating the influence of agent traits on the diffusion process. Our findings emphasize the advantages of LLM-based simulations over traditional techniques, as they uncover underlying causes of information spread---such as agents promoting discussions---beyond the predefined rules typically employed in existing agent-based models. Additionally, we evaluate three countermeasure strategies, discovering that brute-force blocking influential agents in the network or announcing news accuracy can effectively mitigate misinformation. However, their effectiveness is influenced by the network structure, highlighting the importance of considering network structure in the development of future misinformation countermeasures. 
\end{abstract}


\pagestyle{fancy}
\fancyhead{}


\maketitle 


\section{Introduction}
In today's digital age, social platforms play a pivotal role in the diffusion of information, enabling the rapid spread of content through user interactions. However, this ease of sharing also facilitates the swift propagation of misinformation. Here, misinformation is defined as false or misleading information that is spread either intentionally or unintentionally \cite{zareie2021minimizing}. The widespread circulation of misinformation across social networks poses a significant threat, undermining public trust, distorting democratic processes, and causing extensive societal harm. Therefore, understanding the underlying mechanisms of misinformation diffusion is crucial for developing effective strategies to mitigate its impacts and preserve the integrity of information ecosystems.

Information diffusion can be studied from both macro- and micro-level perspectives. Macro-level approaches focus on population-based modeling of large-scale dissemination patterns, such as the dynamics of information cascades in large communities. In contrast, micro-level analyses assess the likelihood of individual users engaging with and propagating information through actions like retweets, shares, or likes \cite{liu2023survey}. Current methodologies, including surveys on user behavior and agent-based models simulating interactions among autonomous agents, have been widely employed \cite{burbach2019shares, liu2023survey}. However, these traditional methods often fall short in capturing the complex network structures that enable the emergence of collective behavior. For example, surveys may suffer from self-report bias, leading to inaccurate representations of user behavior. Additionally, conventional agent-based models often rely on mathematical assumptions, such as predefined parameter values and functional forms, that may not fully reflect real-world dynamics. These models typically necessitate the specification of functions that govern the probability of agents forwarding or forgetting a piece of information \cite{sulis2020simulation}. While these assumptions allow for simplified simulations, they can limit the model’s ability to capture more nuanced behaviors, such as the influence of diverse user motivations or the intricate dynamic of network interactions. 

To address these shortcomings, this work introduces the use of LLMs in multi-agent simulations to study the diffusion of fake news across various social network typologies. LLM-driven simulations bridge both individual-level (micro) and population-level (macro) analyses, enabling the prediction of trends across populations over time as well as the detailed examination of individual interactions. Unlike traditional models, LLM-powered multi-agent simulations are less data-intensive and highly adaptable to the complexities of evolving social networks \cite{gong2007research, liu2023survey}. This adaptability allows for the exploration of diverse scenarios and interventions without real-world risks. By simulating interactions with greater detail, LLMs provide deeper insights into the mechanisms of fake news propagation across different network structures. 

The contributions of this work can be summarized as follows: 
\vspace{-1em}
\begin{itemize}
\item We present a novel LLM-driven multi-agent simulation framework specifically designed to study fake news diffusion across different network structures. This approach offers a unique and advanced method for simulating complex interactions within social networks without relying on predefined assumptions.
\item We systematically model the influence of network typology and agent personalities (e.g., Big Five personality traits) on the spread of fake news to validate the effectiveness of the LLM-driven simulation. 
\item We investigate the interplay between network configurations and agent characteristics to assess various intervention strategies aimed at curbing fake news diffusion. Our finding highlight that the effectiveness of strategies, including blocking influencers, and implementing accuracy checks, significantly depends on the underlying network structure, emphasizing the need for tailored approaches in misinformation countermeasures. 
\end{itemize}



\section{Related Work}

\textbf{Information Diffusion.} Numerous studies have investigated how misinformation spreads, often using simulated networks that mimic real-world structures. Information dissemination models are typically categorized into explanatory models---such as epidemic, improved, network structure, and competitive---and predictive models. These models help capture dynamic behaviors within social networks, providing insights into how information propagates \cite{liu2023survey}.

Agent-based simulations have been widely employed to study misinformation diffusion in social networks. For instance, Burbach \etal\ \cite{burbach2019shares} applied agent-based modeling to investigate how individuals behave within online social networks, investigating how the characteristics of interconnected users contribute to the spread of both factual and false information. 
Gong \etal\ \cite{gong2007research} conducted multi-agent simulations on epidemic news, examining the influence of social relationships, trust in news, and network dynamics on the rate of information spread. They concluded that reducing the number of social relationships could effectively mitigate panic. While these studies highlight the potential of simulations to inform better countermeasure strategies, agent-based simulations still rely on certain assumptions, such as the distribution of news propagation over time. Moreover, these existing studies do not examine the impact of different network structures on information diffusion as we conduct in this study. 

Burbach \etal\ \cite{burbach2020opinion} conducted a study to investigate the factors influencing whether users disseminate information using an agent-based model. They varied content types, network structures, and incorporated a personality model. However, their research did not focus specifically on fake news or explore countermeasures against misinformation. Additionally, the agent-based model they applied relied on predefined rules and did not fully capture the complexities of human behavior or how network structures interact with these factors. In contrast, our study employs LLMs to simulate more nuanced agent behaviors in the context of fake news propagation and countermeasure efficacy.

\textbf{Misinformation Countermeasures.} Countermeasures against misinformation are essential to prevent its harmful impact on public opinion. Gausen \etal\ \cite{gausen2021can}
demonstrated that agent-based modeling is useful for evaluating strategies to combat misinformation. They examined three interventions: rule-based policies which work by blocking users and removing posts based on complaints; societal inoculation which prepares individuals to resist misinformation by building mental antibodies; and accuracy flags, which shifts users' attention by verifying the accuracy of content. They found that inoculation and accuracy flags are effective in minimizng misinformation spread. However, these simulations often rely on pre-defined rules that limit their ability to capture more complex behaviors. Moreover, they did not examine how different network structures might affect the success of these countermeasures. A recent study \cite{liu2024skepticism} has also explored using LLMs to generate persuasive anti-misinformation messages to strengthen public immunity against false information. However, this study focused on changing the attitude of agents, such as transitioning from suspicion to belief. In contrast, our work emphasizes the action of information diffusion rather than the change in attitude.

\textbf{LLM Simulation.} With the rapid advances in LLMs, recent research has explored their potential in simulating the social interactions. For example, Part \etal\ \cite{park2023generative} created a virtual town populated with LLM-powered agents, providing valuable insights into the use of LLMs as a robust tool for agent-based simulation and their potential in modeling social behaviors. Gao \etal\ \cite{gao2023s} introduced a social network simulation system that leverages LLM-powered agents to represent users within a social network. Their work primarily assessed the system's efficiency through case studies on gender discrimination and nuclear energy, demonstrating the versatility of LLMs in simulating social interactions. The use of LLM for social simulation also extends to the news diffusion area. Liu \etal\ \cite{liu2024skepticism} employed LLM-based multi-agent simulations to investigate the spread of fake news, with a focus on evolving attitudes of agents toward misinformation and the effectiveness of intervention strategies. Similar to these preceding work, our work also explores the potential of LLMs in simulating social issues, in particular news diffusion, but different from the existing work, our study focuses on the diffusion process itself, examining how far misinformation can spread under varying network typologies, offering a complementary perspective to existing work.

\section{LLM Multi-agent Simulation}
The LLM multi-agent simulation framework presented in this work consists of two primary components: the network structure $G$ and the interactions among $N$ LLM-simulated agents, each representing an individual within a social network. These agents engage in interactions via the edges $e_{ij}$ within the network, simulating the connections between agent $i$ and agent $j$, which are typically observed in social media platforms. 

The simulation is conducted over $T$ days, during which agents interact with their connected peers using decision-making processes powered by LLMs. At the onset of the simulation, only one agent---referred to as the source agent---receives the news, initiating the process of news diffusion. This setup enables an analysis of how information spreads through the network over time, offering insights into the dynamics of news propagation across various network structures. 

\subsection{Agent Simulation}
This subsection describes the LLM-simulated agent profiles and the statuses that agents may possess during news diffusion.

\subsubsection{Agent Profile}
Consistent with prior research, each agent is randomly assigned a persona that encompasses characteristics, including age, gender, and the Big Five personality traits \cite{burbach2019shares, liu2024skepticism}.

\textbf{Gender:} Each agent has a 50\% probability of being female.

\textbf{Age:} The ages of simulated agents are drawn from a gamma distribution with a mean of 28.5 and a standard deviation of 9.54, as the distribution of age is often positively skewed. Subsequently, the age is rounded to the nearest integer. To ensure the reasonableness of our simulated data, we adhere to statistical mean and standard deviation from surveys conducted in previous studies \cite{burbach2019shares} and the mean is also close to the average age of social media users as reported online \cite{website}. 

\textbf{Big Five:} The Big Five model offers a comprehensive framework for evaluating personality traits across five dimensions: openness, conscientiousness, extraversion, agreeableness, and neuroticism. This model has been widely applied in numerous studies to characterize individuals' personalities, including in existing LLM simulated agent frameworks for attitude dynamics toward fake news \cite{burbach2019shares, liu2024skepticism}. Our approach simulates the trait scores using a multinormal distribution based on the provided statistics from the work \cite{burbach2019shares} in Table \ref{tab:statistics}, and categorizes each trait into high and low levels using the 50th percentile.

\begin{table}[h]
    \centering
    \caption{The Big Five trait scores are sampled using a multinormal distribution with the given mean, standard deviation and the significant correlation scores.}
    \begin{tabularx}{\columnwidth}{lXX} 
        \hline
         & \textbf{Mean} & \textbf{Std}  \\
        \hline
        Extraversion & 4.02 & 1.18 \\
        Agreeableness & 3.81 & 0.89 \\
        Conscientiousness & 4.14 & 0.99 \\
        Neuroticism & 3.43 & 1.12 \\
        Openness & 4.52  & 1.07 \\
        \hline
        \textbf{Significant correlations}  \\
        \hline
        Extraversion vs. Agreeableness & 0.184 \\
        Extraversion vs. Neuroticism &-0.236 \\
        \hline
    \end{tabularx}
    \label{tab:statistics}
\end{table}

\subsubsection{Agent Status: }
In the network, each agent possess a status indicating their role within it. We categorize agents into one of three statuses at the end of day $t$: the set of spreaders, denoted as $S_t$, includes agents who have received the news and will propagate it to others on the following day. Dead ends, represented as $D_t$, are agents who have received the news but choose not to disseminate it further. Reached individuals, denoted as $R_t$, encompass all agents who have received the news throughout the simulation, including both spreaders and dead ends. The remaining agents are classified as unreached, meaning they have not encountered the news at time $t$ during the simulation. 

\subsection{Network Simulation}
To study the influence of network structures on information diffusion, this work focuses on the following well-known network structures in the literature. In all these simulated networks, the edges are unweighted and undirected, signifying a mutual relationship between each pair of agents, which constitutes the basis for social influence \cite{hsiao2021evaluating, burbach2020opinion}.

\textbf{Random Network:} In a random network, each possible edge $e_{ij}$ is included in the network with a fixed probability, independently of every other edge. This network is useful for studying baseline behavior and comparing more complex network structures. 

\textbf{Scale-free Network:} A scale-free network is characterized by a degree distribution that follows a power law. This means a few nodes have a very high degree (i.e., influencers), while most nodes have a relatively low degree (i.e., ordinary social media users), exemplifying the "rich get richer" phenomenon, where highly connected agents are more likely to attract new connections, leading to the emergence of influential hubs within the network. 

\textbf{High Brokerage Network:} A high-brokerage network refers to a structure where certain nodes, known as brokers, play a critical role in connecting different parts of the network. These brokers facilitate interactions between otherwise disconnected subgroups, controlling the flow of information across the network.

 Details of the network statistics are shown in Table~\ref{tab:network_statistics}. The slight difference in the number of agents between the scale-free network and other networks is implemented to ensure comparable network density across simulations with different structures. Since our primary focus is on the proportion of agents who forward and receive the news, rather than the absolute number of agents involved, this variation does not impact the analysis of the results.
 
 Network density is the ratio of the actual number of edges $E$ in the network to the total possible number of edges in a fully connected network $\frac{2E}{N(N-1)}$. Degree refers to the number of edges connected to each node. Average path length measures the average number of steps it takes to get from one node to another across the network, calculated as $\frac{\sum_{i \neq j}d(i, j)}{N(N - 1)}$, where $d(i,j)$ is the shortest path distance between nodes $i$ and $j$. This provides an indication of how separated or well-connected the network is. The average clustering coefficient measures the extent to which nodes in a graph tend to cluster together, with the clustering coefficient for a node $i$ defined as $\frac{2e_i}{k_i(k_i - 1)}$, where $e_i$ is the number of edges between the neighbors of node $i$, and $k_i$ is the degree of node $i$. Modularity measures the strength of division of a network into clusters, calculated as $\frac{1}{2E}\sum_{i,j}[A_{ij} - \frac{k_i k_j}{2E} \delta(c_i, c_j)]$, where $A_{ij}$ is the adjacency matrix, $k_i$ is the degree of node $i$ and $\delta(c_i, c_j)$ is 1 if nodes $i$ and $j$ belong to the same community, and 0 otherwise.

\subsection{Agent Interaction Simulation}
At the beginning of the simulation, every agent is given a persona and starts with the status of "unreached," except for the source agent, who initiates the news diffusion.

\begin{figure}
    \centering
    \includegraphics[width=0.45\textwidth]{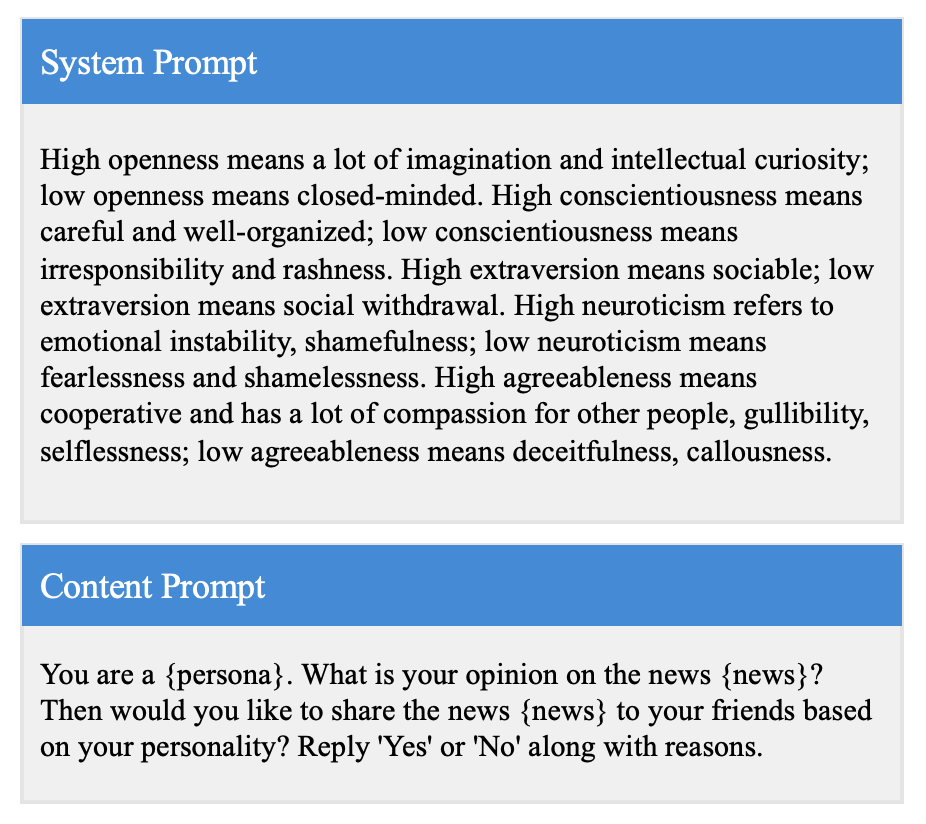} 
    \caption{The prompt designed for LLM agents to make decisions on news diffusion without any intervention strategy. }
    \label{fig:prompt}
\end{figure}

Throughout each day of the simulation, agents who received the news the previous day (those in the set $R_{t - 1}$) must decide whether or not to share the news with their friends. This decision-making process is guided by prompts influenced by their assigned personality traits, as shown in Figure~\ref{fig:prompt}. These prompts are designed to simulate how personality traits affect the likelihood of spreading the news.

If an agent $i$ chooses to share the news, their status changes to "spreader" for that day, and their neighbors or friends in the network will receive the news. These newly informed neighbors are then added to the set of agents $R_{t}$ for the next day. Once an agent has shared the news, they will not share it again in later rounds of the simulation.

\begin{table*}[ht]
    \centering
    \caption{The network statistics of different network structures. Each network structure is generated with varying random seeds, which are kept consistent across the different network structures.}
    \begin{tabularx}{\textwidth}{llllllll} 
        \hline \textbf{Network} & \textbf{Agents}
         & \textbf{Density} & \textbf{Mean Degree} & \textbf{SD Degree}& \textbf{Ave Path}& \textbf{Ave Cluster }& \textbf{Modularity} \\
        \hline
        Random Network & 300 & $0.08\pm0.002$ &  $12.072\pm0.210$ & $3.3687\pm0.128$ & $2.5633\pm0.014$ & $0.0402\pm0.002$ & $0.2585\pm0.005$ \\
    
        Scale-free Network & 288 & $0.08\pm0$ &  $11.75\pm0$ & $9.5395\pm0.416$ &$2.4758\pm0.011$ & $0.1086\pm0.007$ &  $0.2456\pm0.005$\\
        High Brokerage Network & 300 & $0.08\pm0$  & $11.876\pm 0.116$ & $1.6816\pm0.11$5 & $2.9801\pm0.033$ & $0.6002\pm0.011$ & $0.7245\pm0.007$ \\
        \hline
    \end{tabularx}
    \label{tab:network_statistics}
\end{table*}

\section{Experiments}
\subsection{Research Objectives}
This study aims to explore the potential of using LLMs to simulate the diffusion of misinformation across different social network structures. We treat personality as a control variable, focusing on how varying network structures---such as random, scale-free, and high-brokerage networks---impact the spread of misinformation. Our investigation highlights the advantages of LLM-based simulations over traditional agent-based modeling, particularly in capturing complex agent interactions and behaviors without defining rules governing how agents behave. Additionally, we examine how the effectiveness of countermeasures to curb misinformation varies depending on the underlying network structure.

\subsection{Implementation Details}
The language model used in our simulation is \textit{GPT-Turbo-3.5-1106}, with the temperature set to 0 for reproducibility. We define both the agents and the simulated network using the Python library Mesa. The source agent is the one with the highest connectivity, and the simulation is run over $T = 7$ days, as we observe that the diffusion process typically converges within this time frame. To ensure consistency and comparability across different network structures, we employ the same random seed for all network types. Within simulations of the same network type, we introduce variations in seed values and randomly select 5 different fake political news from the FakeNewsNet dataset\footnote{https://github.com/KaiDMML/FakeNewsNet/tree/master}, which contains information from news content, social context, and spatiotemporal data as described in \cite{shu2020fakenewsnet}. These selected fake news articles remain consistent across scenarios. This allows us to maintain uniformity in the analysis while comparing diffusion patterns across different network typologies.

To visually represent the propagation of news within the network over time, we track the proportion of agents who have received the news by the end of each day $\frac{|R_t|}{N}$ and the proportion of agents who have forwarded the news by the end of each day $\frac{|S_t|}{N}$, with the assumption that agents who forwarded the news were somewhat influenced.


\subsection{LLM Simulation Results} \label{sec:experimental_results}
\subsubsection{The influence of personality on news diffusion}
We begin by exploring how the personalities of agents influence their decision to share news. Specifically, we focus on identifying which of the Big Five personality traits make an agent more prone to disseminating fake news. To investigate this, we run simulations using a random network structure. In each simulation, one trait from the Big Five model is set to either a high or low value for all agents, while the other traits are randomly assigned.

\begin{figure*}
    \centering
    \includegraphics[width=\textwidth]{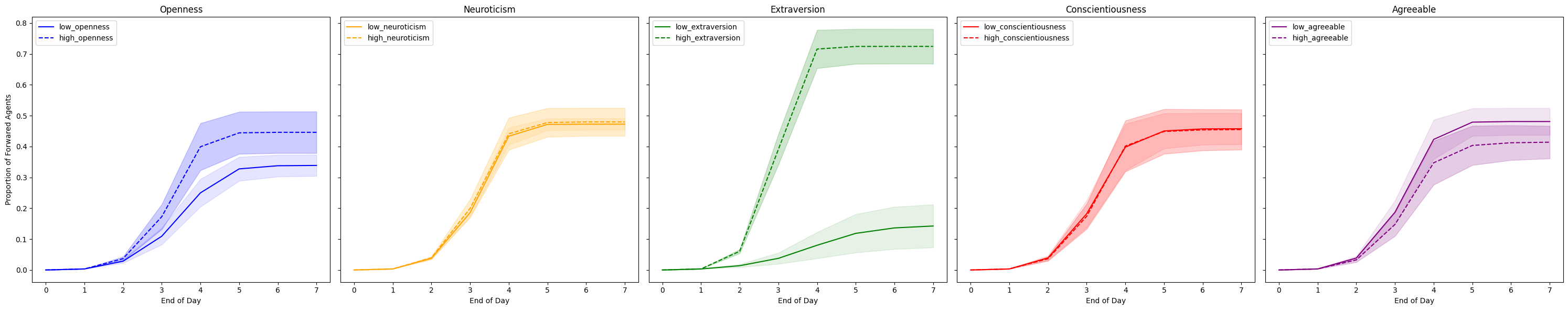} 
    \caption{The influence of agents' personalities on news diffusion within the random network.}
    \label{fig:personality}
\end{figure*}

Figure~\ref{fig:personality} depicts the news diffusion patterns among agents with different personality traits. It tracks how the proportion of agents sharing the news changes as we modify a single personality trait in the LLM simulations. The findings highlight that Extraversion and Openness play a crucial role in the diffusion process. Agents exhibiting high levels of these traits were significantly more likely to share the news compared to those with lower levels. This supports earlier research showing that individuals high in extraversion and openness are more inclined to perceive misinformation as credible and subsequently share it \cite{sampat2022fake, ahmed2022social}.

The consistency between the LLM-simulated agents' decisions to share news or not, as compared to existing work, suggests that LLM-based simulations could serve as a reliable method for studying the spread of information. Additionally, this demonstrates the value of using LLM-based simulations to model how agents with specific personality traits make sharing decisions, offering a distinct advantage over traditional survey methods, which rely on self-reported behavior from participants.

\begin{figure}
    \centering
\includegraphics[width=0.45\textwidth]{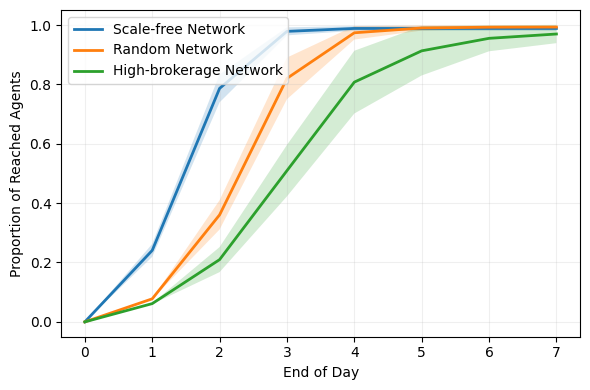} 
    \caption{The proportion of agents who have received the news under different network structures.}
    \label{fig:reached}
\end{figure}

\subsubsection{The influence of network structure}
During the simulations, the initial source agent sometimes chooses not to forward the news, halting the diffusion process from the start. To ensure we focus on effective news diffusion, we only analyzed simulations where the source agent does forward the information.

Figure~\ref{fig:reached} shows how the structure of the network affects the proportion of agents who received the news, known as the ``reached agents.'' Our results show that news spreads most rapidly in scale-free networks, followed by random networks, with the slowest spread occurring in high-brokerage network.  This pattern is logical, as scale-free networks, despite having a similar average number of connections as other networks, contain a few highly connected agents who facilitate quick diffusion. In contrast, high-brokerage networks experience rapid initial spread within tightly connected clusters but struggle to disseminate the news across other clusters, limiting its overall reach. To confirm these differences, we employ the Wilcoxon rank-sum test on the diffusion rate, measuring how quickly at least 50\% of agents received the news. The results, showing p-values lower than 0.05 for all network pair comparisons, confirm that the dynamics of diffusion are significantly affected by the network structure \cite{centola2010spread}.

This demonstrates that the LLM simulation can produce realistic diffusion patterns based on behavior of simulated agents, reducing the need for strict predefined rules typically required in traditional models.


\begin{figure}
    \centering
    \includegraphics[width=0.5\textwidth]{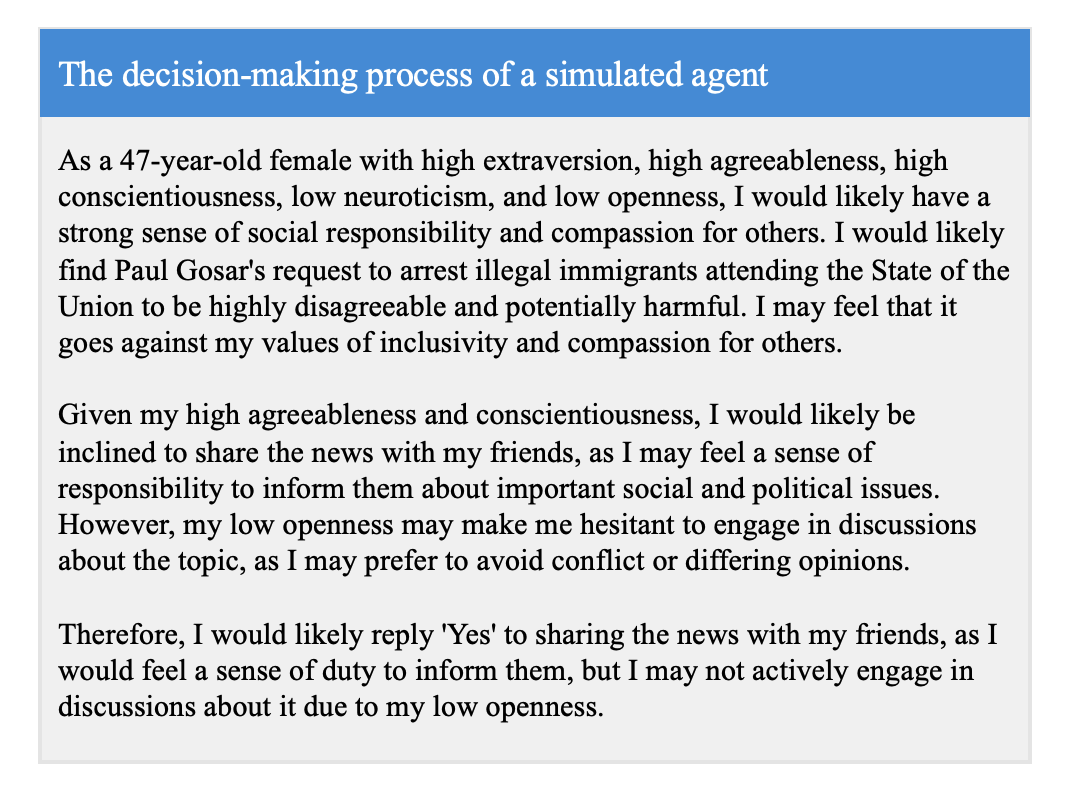}  
    \caption{A micro-level observation reveals the details of an agent's thought process.}
    \label{fig:persona}
\end{figure}

\begin{figure}
    \centering
    \includegraphics[width=0.5\textwidth]{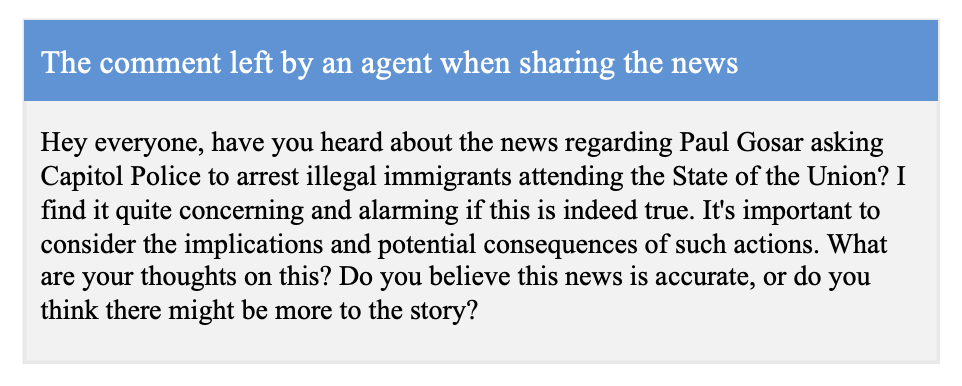}  
    \caption{An example showing the comment left by an agent when sharing the news.}
    \label{fig:comment_exp}
\end{figure}

\begin{figure*}
    \centering
    \includegraphics[width=\textwidth]{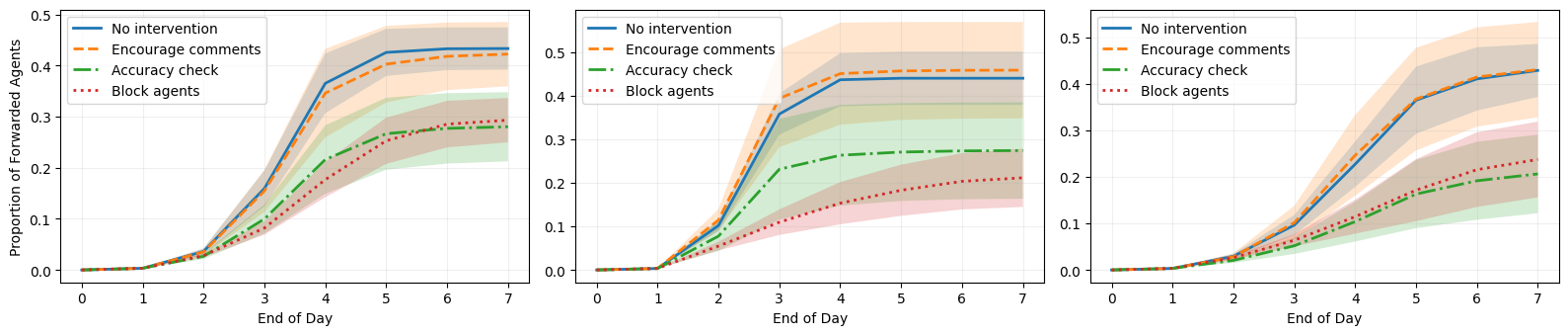}  
    \caption{Comparison of countermeasures across different network structures based on the proportion of agents sharing the news. Left: Random network, Middle: Scale-free network, Right: High-brokerage network.}
    \label{fig:countermeasure_forward}
\end{figure*}

\begin{figure*}
    \centering
    \includegraphics[width=\textwidth]{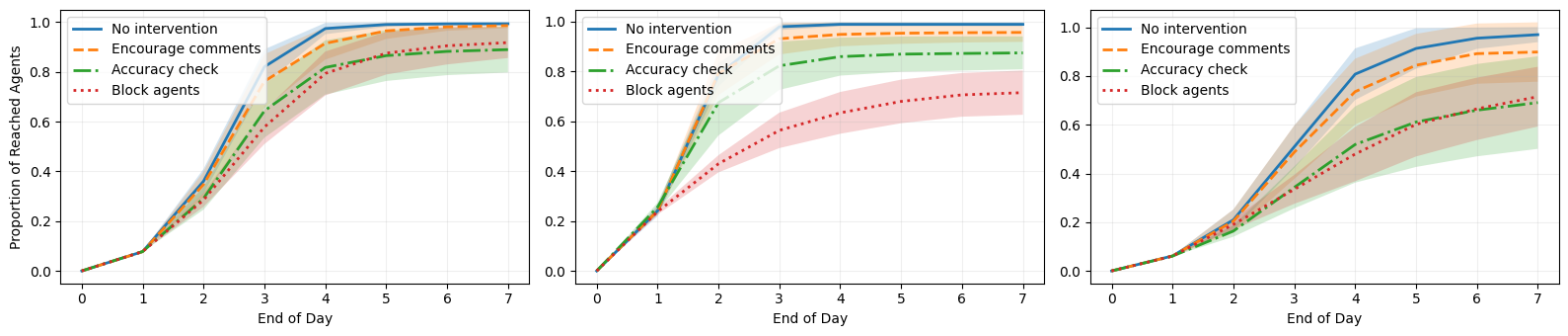}  
    \caption{Comparison of countermeasures across different network structures based on the proportion of agents who have received the news. Left: Random network, Middle: Scale-free network, Right: High-brokerage network.}
    \label{fig:countermeasure_reached}
\end{figure*}

\subsection{Fake News Intervention}
\subsubsection{Micro-level Observation} 
Before exploring countermeasures across different network structures, we first highlight a key advantage of using LLM-based simulations: the ability to provide detailed micro-level observations of individual agents. Unlike traditional agent-based models that rely on predefined rules or probability-based actions, LLM-based simulations allow for a more explicit understanding of each agent’s decision-making process. This is achieved by interrogating the rationale behind their actions, providing insights into the specific motivations driving behavior.

For example, instead of simply assuming an agent will forward news based on a fixed probability, LLM simulations model how individual personality traits, experiences, or contextual factors influence the decision. Figure~\ref{fig:persona} illustrates the behavior of an agent with high extraversion, agreeableness, and conscientiousness, combined with low neuroticism and openness. These traits directly shape the agent’s choice to propagate news, showcasing how LLM-based models can reveal the thought processes behind actions.

\subsubsection{Intervention Strategies}
Section~\ref{sec:experimental_results} demonstrates that news diffusion is influenced by both agents' personalities within the network and the structure of the network. Building on these findings, this section leverages the LLM-driven multi-agent simulation to explore countermeasures aimed at curbing fake news diffusion. Specifically, we compare the following strategies: 

\begin{figure}
    \centering
    \includegraphics[width=0.5\textwidth]{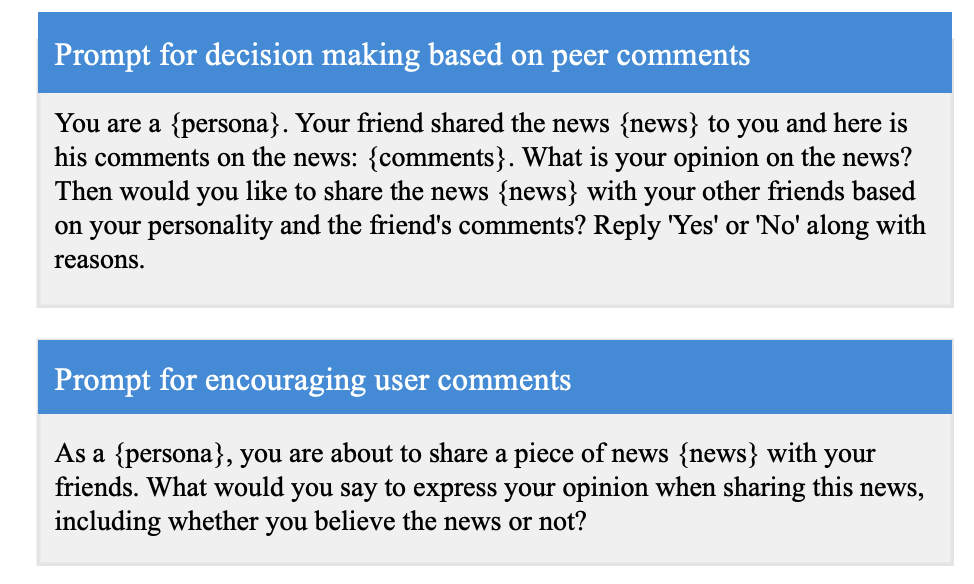}  
    \caption{Prompts for agents' decision-making based on peer comments and requesting agents' comments when sharing the news.}
    \label{fig:comment_prompts}
\end{figure}

\textbf{Encouraging Commenting on Shared News:} One advantage of LLM-based simulations is the ability to observe the micro-level decision-making process of agents. We notice that agents might choose to forward news not only to disseminate it but also to foster discussion or engagement within their networks. With this in mind, we explore the hypothesis that encouraging agents to add comments when sharing news could potentially reduce the spread of misinformation. The designed prompts are shown in Figure~\ref{fig:comment_prompts}. The rationale behind this intervention is that requiring agents to articulate their thoughts may prompt them to think critically about the content they are about to share, thereby reducing impulsive sharing behavior.


\textbf{Announcing the Accuracy of Shared News:} Fact-checking algorithms are commonly used to verify the veracity of news; however, they might confirm misinformation after it has already been disseminated for some time \cite{shu2017fake}. To simulate this scenario, we test an intervention strategy in which we assume that the veracity of news is confirmed once a certain threshold of spread is reached.

In our simulated high-brokerage network setup, where 10\% of agents form a highly connected hub, we hypothesize that when at least 10\% of agents have received the news, the information will be officially refuted if it is false. Consequently, agents who have received the news will be informed of its inaccuracy during their decision-making process regarding whether to share it further. We aim to examine how this intervention strategy influences information diffusion across different network structures.

\textbf{Blocking a Percentage of Influencers:} This strategy focuses on blocking key influencers in the network to curb the spread of misinformation. Influencers can be identified through various measures of centrality in a social network \cite{huynh2019some}. Since previous experiments showed that agents with high levels of openness and extraversion are more prone to sharing news, we target these traits in our intervention strategy. Specifically, we identify the top 20\% of agents who score high on openness or extraversion and also have the most connections (referred to as the highest degree centrality in network research). Regulatory bodies are assumed to detect the spread of misinformation over time and initiate the process of blocking these high-influence agents. To maintain consistency with the threshold set in the accuracy-check strategy, this blocking intervention is triggered once 10\% of agents in the network have received the news. If any of these highly connected agents have not yet been exposed to the news by this point, they are preemptively blocked to prevent further spread of misinformation.

\subsubsection{Intervention Results}
The simulation results are presented in Figure~\ref{fig:countermeasure_forward}
and Figure~\ref{fig:countermeasure_reached}. Wilcoxon rank sum test is utilized to evaluate the statistical significance of differences in terms of the ultimate proportion of forwarded agents and the reached agents. From these figures, we observe that asking agents to comment on shared news does not significantly foster critical thinking among other agents (p-values greater than 0.05), thus failing to halt the spread of misinformation. In some cases, agents share the news with the intent to verify its accuracy within their social circles and encourage discussion, as shown in Figure~\ref{fig:comment_exp}. This behavior indicates that merely encouraging agents to comment while sharing news does not effectively reduce the dissemination of fake news across all simulated network structures.

We also observe that both blocking influencers and announcing the accuracy of the news significantly reduce the proportion of agents who forward the news across all three network structures, compared to the strategy of asking agents to comment when sharing. Specifically, blocking influencers is significantly more effective (p-values < 0.05) than no interventions across all network structures. On the other hand, announcing the accuracy of the news proves effective only in random and high-brokerage networks (p-values < 0.05), but not in scale-free networks. This suggests that different network structures respond differently to interventions, highlighting the need for tailored strategies when attempting to curb misinformation spread.

For the proportion of agents who ultimately receive the news, both blocking influencers and announcing the accuracy of the news outperform the strategies of commenting and no interventions across all network structures. However, in scale-free networks, blocking is significantly more effective than the accuracy-check intervention (p-values < 0.05). 

Even though there is no significant difference between blocking influencers and accuracy checks within random and high-brokerage networks (p-values greater than 0.05), both strategies tend to be more effective in high-brokerage networks compared to random ones. This is likely due to the structural characteristics of high-brokerage networks, where narrow bridges between clusters create bottlenecks that slow down the diffusion of news. Since the countermeasures are applied after 10\% of the agents have been exposed, it means that entire clusters may already be affected, especially since the source agent is the node with the highest degree. As a result, interventions become more impactful in high-brokerage networks where the spread is naturally constrained by the network structure, whereas in random networks, news tends to spread more uniformly across the network.


If we aim to achieve similar effectiveness in a random network as seen in the high-brokerage network, earlier interventions may be required. Since random networks lack the clear structural bottlenecks present in high-brokerage networks, news can spread more uniformly, and delaying the intervention could allow the news to reach a larger portion of the network. This observation is important because, while random networks can occur in real life, social media platforms often exhibit a scale-free structure, particularly in follower-followee dynamics, as supported by previous research \cite{myers2014information}. In scale-free networks, a few highly connected nodes (influencers) drive the spread of information, making targeted interventions more effective. In contrast, random networks, which lack these influential hubs, require broader and earlier interventions to contain misinformation before it spreads widely.

These observations show that different network structures respond differently to interventions, highlighting the need for tailored strategies when attempting to curb misinformation spread. While both strategies can curb the spread of misinformation, blocking is particularly crucial in network structures where highly connected individuals play a dominant role in facilitating news diffusion.

\vspace{-0.4em}

\section{Conclusion}
In this study, we leverage LLM-driven simulations to explore the dynamics of misinformation diffusion across diverse social network structures. Rather than aiming for a perfect simulation of news dissemination, our focus is on demonstrating the advantages of LLM-based simulations over traditional agent-based models and survey studies. While user behavior may sometimes correlated with their beliefs (e.g., whether they trust a news), this work emphasizes agents' actions rather than their attitudes. 

By simulating agent interactions with varying personality traits and network typologies, our results reveal that personality traits, particularly high extraversion and openness, would increase the spread of news, contributing to more realistic simulations compared to traditional agent-based models.

We examine the impact of three countermeasures---encouraging comments on shared news, announcing the accuracy of the news, and blocking influential agents---across random, scale-free, and high-brokerage networks. While encouraging comments does not significantly reduce the spread of misinformation, announcing accuracy and blocking influencers proved to be effective strategies, particularly in scale-free networks. In high-brokerage and random networks, both methods performed similarly, though earlier interventions were suggested to be more impactful in random networks due to their more uniform structure.

These findings demonstrate the potential of LLM-based simulations to provide deeper insights into misinformation spread and emphasize the importance of tailoring countermeasures to specific network characteristics. Our research opens up new possibilities for using LLMs to simulate complex social behaviors and devise effective strategies to combat misinformation.




\section{Limitations and Future Directions}
While this study provides valuable insights into the use of LLM-based simulations for modeling misinformation diffusion, several limitations should be noted. First, the simulations rely on a simplified representation of personality traits and decision-making processes, which may not fully capture the complexities of human behavior. Real-world decisions are influenced by a wider range of factors, including emotions, personal experiences, and external pressures. Additionally, real-world actions extend beyond simple sharing and can include behaviors such as liking, reporting content, and more, which were not incorporated into our model. Future work could explore the integration of more sophisticated psychological and sociocultural factors to enhance the agent interactions and better capture the complexities of misinformation diffusion.

Second, while we explore the effects of countermeasures like blocking influencers and accuracy-checking, these interventions were applied in a relatively simplistic manner. Real-world interventions often involve more gradual, adaptive, and multifaceted approaches. Future work could simulate more dynamic and context-sensitive countermeasures, potentially incorporating strategies like gradual influencer demotion or personalized accuracy notifications.

Furthermore, in reality, an individual's network location is not exogenous; it may be influenced by personality traits, age, or gender. However, our model assumes that these factors are independent of one another. Additionally, information diffusion—particularly the spread of fake news on social media—is complicated by bot activities, as social bots are intentionally created to manipulate public opinion and alter the trajectory of information flow. This aspect has not been accounted for in this study.
\bibliographystyle{ACM-Reference-Format}
\bibliography{ref}



\end{document}